\documentclass[aps,preprint, floatfix]{revtex4}

\usepackage{graphicx,color}
\usepackage{psfrag}
\usepackage{amssymb}
\usepackage{eso-pic}
\usepackage[dvips,letterpaper]{geometry} 
\usepackage{amsmath}

\begin{document}

%
%
%
%
\def\oti{{\otimes}}
\def\lb{ \left[ }
\def\rb{ \right]  }
\def\tilde{\widetilde}
\def\bar{\overline}
\def\hat{\widehat}
\def\*{\star}
\def\[{\left[}
\def\]{\right]}
\def\({\left(}		\def\BL{\Bigr(}
\def\){\right)}		\def\BR{\Bigr)}
	\def\BBL{\lb}
	\def\BBR{\rb}
%
%
\def\zb{{\bar{z} }}
\def\zbar{{\bar{z} }}
\def\frac#1#2{{#1 \over #2}}
\def\inv#1{{1 \over #1}}
\def\half{{1 \over 2}}
\def\d{\partial}
\def\der#1{{\partial \over \partial #1}}
\def\dd#1#2{{\partial #1 \over \partial #2}}
\def\vev#1{\langle #1 \rangle}
\def\ket#1{ | #1 \rangle}
\def\rvac{\hbox{$\vert 0\rangle$}}
\def\lvac{\hbox{$\langle 0 \vert $}}
\def\2pi{\hbox{$2\pi i$}}
\def\e#1{{\rm e}^{^{\textstyle #1}}}
\def\grad#1{\,\nabla\!_{{#1}}\,}
\def\dsl{\raise.15ex\hbox{/}\kern-.57em\partial}
\def\Dsl{\,\raise.15ex\hbox{/}\mkern-.13.5mu D}
%
%
\def\ga{\gamma}		\def\Ga{\Gamma}
\def\be{\beta}
\def\al{\alpha}
\def\ep{\epsilon}
\def\vep{\varepsilon}
\def\la{\lambda}	\def\La{\Lambda}
\def\de{\delta}		\def\De{\Delta}
\def\om{\omega}		\def\Om{\Omega}
\def\sig{\sigma}	\def\Sig{\Sigma}
\def\vphi{\varphi}

%
%
\def\CA{{\cal A}}	\def\CB{{\cal B}}	\def\CC{{\cal C}}
\def\CD{{\cal D}}	\def\CE{{\cal E}}	\def\CF{{\cal F}}
\def\CG{{\cal G}}	\def\CH{{\cal H}}	\def\CI{{\cal J}}
\def\CJ{{\cal J}}	\def\CK{{\cal K}}	\def\CL{{\cal L}}
\def\CM{{\cal M}}	\def\CN{{\cal N}}	\def\CO{{\cal O}}
\def\CP{{\cal P}}	\def\CQ{{\cal Q}}	\def\CR{{\cal R}}
\def\CS{{\cal S}}	\def\CT{{\cal T}}	\def\CU{{\cal U}}
\def\CV{{\cal V}}	\def\CW{{\cal W}}	\def\CX{{\cal X}}
\def\CY{{\cal Y}}	\def\CZ{{\cal Z}}

\def\rvac{\hbox{$\vert 0\rangle$}}
\def\lvac{\hbox{$\langle 0 \vert $}}
\def\comm#1#2{ \BBL\ #1\ ,\ #2 \BBR }
\def\2pi{\hbox{$2\pi i$}}
\def\e#1{{\rm e}^{^{\textstyle #1}}}
\def\grad#1{\,\nabla\!_{{#1}}\,}
\def\dsl{\raise.15ex\hbox{/}\kern-.57em\partial}
\def\Dsl{\,\raise.15ex\hbox{/}\mkern-.13.5mu D}
%
%
%
\font\numbers=cmss12
\font\upright=cmu10 scaled\magstep1
\def\stroke{\vrule height8pt width0.4pt depth-0.1pt}
\def\topfleck{\vrule height8pt width0.5pt depth-5.9pt}
\def\botfleck{\vrule height2pt width0.5pt depth0.1pt}
\def\Zmath{\vcenter{\hbox{\numbers\rlap{\rlap{Z}\kern
0.8pt\topfleck}\kern 2.2pt
                   \rlap Z\kern 6pt\botfleck\kern 1pt}}}
\def\Qmath{\vcenter{\hbox{\upright\rlap{\rlap{Q}\kern
                   3.8pt\stroke}\phantom{Q}}}}
\def\Nmath{\vcenter{\hbox{\upright\rlap{I}\kern 1.7pt N}}}
\def\Cmath{\vcenter{\hbox{\upright\rlap{\rlap{C}\kern
                   3.8pt\stroke}\phantom{C}}}}
\def\Rmath{\vcenter{\hbox{\upright\rlap{I}\kern 1.7pt R}}}
\def\Z{\ifmmode\Zmath\else$\Zmath$\fi}
\def\Q{\ifmmode\Qmath\else$\Qmath$\fi}
\def\N{\ifmmode\Nmath\else$\Nmath$\fi}
\def\C{\ifmmode\Cmath\else$\Cmath$\fi}
\def\R{\ifmmode\Rmath\else$\Rmath$\fi}

\def\barray{\begin{eqnarray}}
\def\earray{\end{eqnarray}}
\def\beq{\begin{equation}}
\def\eeq{\end{equation}}

\def\n{\noindent}

\def\Tr{\rm Tr} 
\def\xvec{{\bf x}}
\def\kvec{{\bf k}}
\def\kvecp{{\bf k'}}
\def\omk{\om{\kvec}} 
\def\dk#1{\frac{d\kvec_{#1}}{(2\pi)^d}}
\def\2pid{(2\pi)^d}
\def\ket#1{|#1 \rangle}
\def\bra#1{\langle #1 |}
\def\vol{V}
\def\adag{a^\dagger}
\def\rme{{\rm e}}
\def\Im{{\rm Im}}
\def\pvec{{\bf p}}
\def\fermiS{\CS_F}
\def\cdag{c^\dagger}
\def\adag{a^\dagger}
\def\bdag{b^\dagger}
\def\vvec{{\bf v}}
\def\muhat{{\hat{\mu}}}
\def\vac{|0\rangle}
\def\pcut{{\Lambda_c}}
\def\chidot{\dot{\chi}}
\def\gradvec{\vec{\nabla}}
\def\psitilde{\tilde{\Psi}}
\def\psibar{\bar{\psi}}
\def\psidag{\psi^\dagger} 
\def\m{m_*}
\def\up{\uparrow}
\def\down{\downarrow}
\def\Qo{Q^{0}}
\def\vbar{\bar{v}}
\def\ubar{\bar{u}}
\def\smallhalf{{\textstyle \inv{2}}}
\def\smallsqrt{{\textstyle \inv{\sqrt{2}}}}
\def\rvec{{\bf r}}
\def\avec{{\bf a}}
\def\pivec{{\vec{\pi}}}
\def\svec{\vec{s}} 
\def\phivec{\vec{\phi}}
\def\daggerc{{\dagger_c}}
\def\Gfour{G^{(4)}}
\def\dim#1{\lbrack\!\lbrack #1 \rbrack\! \rbrack }
\def\qhat{{\hat{q}}}
\def\ghat{{\hat{g}}}
\def\nvec{{\vec{n}}}
\def\bull{$\bullet$}
\def\ghato{{\hat{g}_0}}
\def\r{r}
\def\deltaq{\delta_q}
\def\gcharge{g_q}
\def\gspin{g_s}
\def\deltas{\delta_s}
\def\gQC{g_{AF}} 
\def\ghatqc{\ghat_{AF}}
\def\xqc{x_{AF}}
\def\mhat{\hat{m}}
\def\xup{x_2}
\def\xdown{x_1}
\def\sigmavec{\vec{\sigma}}
\def\xopt{x_{\rm opt}}
\def\Lambdac{{\Lambda_c}}
\def\angstrom{{{\scriptstyle \circ} \atop A}     }
\def\AA{\leavevmode\setbox0=\hbox{h}\dimen0=\ht0 \advance\dimen0 by-1ex\rlap{
\raise.67\dimen0\hbox{\char'27}}A}
\def\ratio{\gamma}
\def\Phivec{{\vec{\Phi}}}
\def\singlet{\chi^- \chi^+} 
\def\mhat{{\hat{m}}}

\def\Im{{\rm Im}}
\def\Re{{\rm Re}}

\def\xstar{x_*}

\def\sech{{\rm sech}}

\def\Li{{\rm Li}}

\def\dim#1{{\rm dim}[#1]}

\def\ep{\epsilon}

\def\free{\CF}

\def\Fhat{\digamma}

\def\ftilde{\tilde{f}}

\def\muphys{\mu_{\rm phys}}

\def\xitilde{\tilde{\xi}}

\def\CI{\mathcal{I}}

\def\nhat{\hat{n}}

\def\ef{\epsilon_F}

\title{On the viscosity to entropy density ratio for unitary Bose and Fermi Gases}
\author{ Andr\'e  LeClair}
\affiliation{Newman Laboratory, Cornell University, Ithaca, NY}

\bigskip\bigskip\bigskip\bigskip

\begin{abstract}
 
We calculate the ratio of the  viscosity to the  entropy density for both Bose and
Fermi gases in the unitary limit using  a  new approach to the quantum statistical mechanics of gases
based on the S-matrix.  In the unitary limit the scattering length diverges and the S-matrix 
equals $-1$.    For the fermion case we obtain $\eta/s > 4.7$ times the proposed lower bound
of $\hbar/4 \pi k_B$ which  came from the AdS/CFT  correspondence 
for gauge theories,  consistent with the most
recent experiments.   For the bosonic case we present evidence that the gas undergoes a phase transition to
a strongly interacting Bose-Einstein condensate, and is a more perfect fluid, with $\eta/s > 1.3$ times the bound.   

\end{abstract}

\maketitle

\section{Introduction}

One  measure of the perfection of a fluid is its viscosity.      
In recent years new insights on such properties came 
from string theory,  more precisely the AdS/CFT correspondence\cite{Maldacena1,Kovtun}.    
This correspondence relates a conformally invariant (scale invariant)   strongly coupled gauge theory
in $d+1$ space-time dimensions to a gravitational dual in one higher spatial dimension.   
By studying black hole solutions in the higher dimensional theory,  one can study
finite temperature and density properties of the lower dimensional quantum field theory.   
In its  original  version,   the conformal quantum field theory is a certain $N=4$ supersymmetric gauge theory,
which is conformally invariant for all couplings since the beta function vanishes.   
Although such a supersymmetric gauge theory does not describe nature as we currently understand it,
the AdS/CFT correspondence is nevertheless very useful for thinking about these difficult problems 
in new ways and inspiring the study of specific properties that were hardly considered before. 
A prominent example is the ratio of the shear viscosity $\eta$ to the entropy density $s$.    
In natural units with $\hbar = k_B = 1$,  it is a dimensionless quantity.      Using the AdS/CFT correspondence it was
found that the supersymmetric gauge theory  had  $\eta/s = 1/4 \pi$.   It was conjectured that this value 
represents a lower bound\cite{Kovtun},  i.e. 
\beq
\label{bound}
\frac{\eta}{s}  \geq  \inv{4 \pi}   \frac{\hbar}{k_B} 
\eeq
The observed $\eta/s$ for the quark-gluon plasma, studied via heavy ion collisions,  is about  $5$ times this bound\cite{Rick}.
Theoretical estimates for the quark-gluon plasma tend to be lower,  but consistent with the bound.

Note that the bound (\ref{bound}) does not depend on the speed of light.   
Thus, the  natural question arises:  ``Do non-relativistic condensed matter systems respect the conjectured bound,
and which class of fluids has the smallest $\eta/s$?''    For the relativistic fluids studied wth the AdS/CFT correspondence,
the scale invariance plays an important role.    It has thus  been proposed that the unitary Fermi gas may
be the most perfect fluid\cite{Massignan,Gelman,Schafer,Rupak},  in part because it is scale invariant.       
In the so-called unitary limit of a quantum Bose or Fermi gas,  the
scattering length $a$ diverges.   This occurs at a fixed point of the 
renormalization group,  thus these systems provide interesting examples of 
interacting,  scale-invariant theories with dynamical exponent $z=2$, 
i.e. non-relativistic.  
They can be realized experimentally by tuning the scattering
length to $\pm \infty$ using a Feshbach resonance.
(See for instance \cite{Experiment1,Experiment2} and references
therein.)   They are also thought to occur at the surface of
neutron stars.  These systems  have  attracted much theoretical 
interest\cite{Leggett,Nozieres,Ohashi,Ho,HoMueller,Astrakharchik,Nussinov,Perali,LeeShafer,Wingate,Bulgac,Drummond,Burovski,Nishida,Nikolic}.

Because of the scale-invariance,  the only length scales in the problem
are based on the density $n^{1/3}$, 
and the thermal wavelength $\lambda_T = \sqrt{2\pi/mT}$.   Equivalently,
the only energy scales are the chemical potential $\mu$ and the temperature $T$.  
The problem is challenging since there is no small parameter to expand in
such as $n a^3$.    Any possible critical point must occur at a 
specific value of $x=\mu/T$.   This can be translated into 
universal values for $n_c \lambda_T^3$, or for fermions 
universal values for $T_c/T_F$ where $\ep_F = k_B T_F$ is the Fermi energy.  
For instance the critical point of an ideal Bose gas is the simplest example,
where $n_c \lambda_T^3 = \zeta (3/2) = 2.61$.

The models considered  are the simplest 
models of non-relativistic bosons or  fermions with quartic
interactions.    The bosonic model is defined by the action
for a complex scalar field $\phi$. 
\beq
\label{bosonaction}
S =  \int d^3 \xvec dt \(  i \phi^\dagger  \d_t \phi - 
\frac{ |\vec{\nabla} \phi |^2}{2m}  - \frac{g}{4} (\phi^\dagger  \phi)^2 \)
\eeq
For fermions, due to the fermionic statistics, 
one needs at least a 2-component field 
$\psi_{\up , \down} $:
\beq
\label{fermionaction}
S = \int d^3 \xvec dt \(  \sum_{\alpha=\up, \down}  
i \psi^\dagger_\alpha \d_t  \psi_\alpha  - 
\frac{|\vec{\nabla}  \psi_\alpha|^2}{2m}   - \frac{g}{2} 
\psi^\dagger_\up \psi_\up \psi^\dagger_\down \psi_\down \) 
\eeq
In both cases,  positive $g$ corresponds to repulsive interactions.   
The bosonic theory only has a $U(1)$ symmetry.   The fermionic theory
on the other hand has the much larger SO(5) symmetry.   
This is evident from the work\cite{Nikolic} which considered 
an $N$-component version with Sp(2N) symmetry,  and noting that
Sp(4) = SO(5).

The original AdS/CFT conjecture was for a specific,  relativistic, and supersymmetric gauge theory.   
There have  been some proposals to use the AdS/CFT correspondence 
to learn about non-relativistic systems\cite{Son,Maldacena,Herzog,Adams}.  
One difficulty is that the conformal symmetry of relativistic systems is 
larger than the Schr\"odinger symmetry of non-relativistic systems,  so the black hole
solutions on the gravity side are less obvious.    Also,   given a  black hole geometry,
it remains unclear which condensed matter system it is dual to.
Thus far,  the AdS/CFT approaches for non-relativistic systems lead to $\eta/s = 1/4 \pi$.
Recent experimental work on the unitary Fermi gas  reports values of $\eta/s$ about 
$4-5$ times the bound\cite{Thomas},   thus it seems unlikely that a gravity dual exists that corresponds exactly to
the unitary Fermi gas.  
Nevertheless,  
it is hoped that one can still discover some general, model-independent  properties, 
in the same way that AdS/CFT for supersymmetric Yang-Mills provided insights into QCD.   

For the remainder of this paper we will describe a novel,  but more conventional approach to 
the problem\cite{PyeTonUnitary}.    The main approximation made is that we only consider 2-body
interactions,  and consistently  resum their contributions to the free energy via an integral equation.
The 2-body interactions are expressed in terms of the zero-temperature S-matrix,  which can be 
calculated exactly.      The viscosity is calculated in a kinetic,  semi-classical approach.  
 The calculation  predicts  a minimum $\eta/s $ of 
$4.7$ times the conjectured lower bound.   This is  consistent with the most recent experiments\cite{Thomas},
which suggests our approximation is a good one.

Theoretical studies have mainly focussed on the fermionic case,   
and for the most part at zero temperature,   which is appropriate for a large Fermi energy.
  The bosonic case has been less studied,  since 
a homogeneous 
bosonic gas with attractive interactions is thought to be unstable against
mechanical collapse, and the collapse occurs before any kind of BEC.  
The situation is actually different for harmonically  trapped gases,  where BEC can occur\cite{trapped}.  
However studies of the homogeneous bosonic case were  based on a small, 
negative scattering length\cite{Stoof,MuellerBaym,Thouless,YuLiLee},
and it is not clear that the conclusions reached  there can be extrapolated 
to the unitary limit.    Since the density of
collapse is proportional to $1/a$\cite{MuellerBaym},   extrapolation to infinite scattering length 
suggests that the gas  collapses at zero density,  which seems unphysical,  since the gas could in
principle be stabilized at finite temperature by thermal pressure.   
 One can also point out that  in the van der Waals gas,  the
collapse is stabilized by a finite size of the atoms,  which renders
the compressibility finite.  In the unitary limit, there is nothing to play
such a role.  
  In the sequel we will present evidence that the unitary 
Bose gas undergoes BEC when $n \lambda_T^3 \approx 1.3$. 
 This lower value compared to the free case 
is consistent with the attractive interactions.   
For this bosonic case,  the minimum $\eta/s$ predicted in \cite{PyeTonUnitary} was only
$1.3$ times the bound,  and  thus may be a better candidate than unitary fermions for the most perfect  strongly 
interacting fluid.

\section{S-matrix,  renormalization group and scattering length}

\def\ghat{\hat{g}}

In this section we describe the  renormalization group fixed point,  the bound state, 
and  the scattering length.   The interplay between all of these properties is most clearly
seen from the S-matrix.    These results are well-known and can be found in the book \cite{book} 
for instance.   We include them here for completeness and in order to consistently describe all 
of our conventions.

The free versions of the models described in the introduction  have a 
scale invariance with dynamical exponent $z=2$, 
i.e. are invariant under
 $$t\to \Lambda^{-2} t, ~
\xvec \to \Lambda^{-1} \xvec$$ 
As we now explain,  the models possess 
 a renormalization 
group fixed point, i.e. quantum critical point, 
where they have 
the same scale invariance.  
The renormalization group behavior can be inferred from the
coupling constant dependence of the S-matrix.     Consider first the  single boson;   the differences for
two-component fermions will be described at the end of this section.   
The S-matrix can be calculated exactly by summing multi-loop ladder diagrams.  
By the Galilean invariane,  
the two-body   S-matrix  depends only on the difference of the incoming momentum of the two particles $\kvec, \kvec'$:  
\beq
\label{3dS}
S(|\kvec - \kvec'|) =   \frac{16 \pi/m g_R  - i  |\kvec - \kvec'|}
{16\pi/ m g_R  + i  |\kvec - \kvec'|}
\eeq
Unitarity of the S-matrix amounts to $S^* S = 1$.  

The momentum space integrals for the higher loop corrections are divergent and an upper cut-off
$\Lambda$ must be introduced.   In the above expression, $g_R$ is the renormalized coupling:
\beq
\label{gR}
\inv{g_R} = \inv{g} + \frac{m \Lambda}{4 \pi^2}
\eeq
   Defining 
$g= \hat{g}/  \Lambda$,  where $\ghat$ is dimensionless,  and requiring $g_R$ to be independent
of $\Lambda$ gives the beta-function:
\beq
\label{betafun}
\frac{ d \ghat}{d \ell}  =  - \ghat - \frac{m}{4\pi^2} \ghat^2 
\eeq
where $\ell =  - \log \Lambda$ is the logarithm of a length scale. 
The above beta function is exact since it was calculated from the exact S-matrix.  
One  thus sees that the theory possesses a fixed point at the negative coupling
$g_* =  -4 \pi^2 / m \Lambda$.

We turn now to the scattering length $a$.   
From the above expression for the S-matrix  one can infer the scattering amplitude and
compute the total cross-section $\sigma$.    Equating 
$\sigma = \pi a^2$
  gives
\beq
\label{a3d}
a (k) =  \frac{m}{2\pi}   \frac{ g_R}{\sqrt{ 1 + (m g_R k /8\pi)^2 }}
\eeq
where here $k$ is the momentum of one of the particles in the center of mass frame.   
If $a(k)$ is measured at very small momentum transfer $|\kvec - \kvec'|
\approx 0$,  this leads
to the definition of the scattering length 
\beq
\label{a3d.2}
a =  \frac{m g_R}{2\pi}  =   \frac{mg}{2\pi (1 -  g/ g_*) } 
\eeq
One sees that scattering length diverges 
at precisely the fixed point $g=g_*$.   Note the  S-matrix becomes $S= -1$.     The scattering length  
$a\to \pm \infty$, depending on from which side $g_*$ is approached.
When $g=g_*^-$, i.e. just less than $g_*$,  then $a\to \infty$, 
whereas when $g=g_*^+$, $a \to -\infty$.

Finally,  we turn to the bound state.    
The S-matrix (\ref{3dS}) has a pole at $k=16 \pi i/mg_R$. 
  Since physical bound states correspond to 
poles at $\Im ( k ) >0$, 
 the bound state exists only for $g$ below $g_*$.
 The energy of this bound state is 
\beq
\label{Ebound}
E_{\rm bound - state} =  -  \frac{128 \pi^2}{m^3 g_R^2}
\eeq
Note that at the fixed point where $g_R$ diverges,   the energy of the bound 
state goes to zero as it should,  since it disappears beyond this point.

Consider now two-component fermions  with the action in the Introduction.   
The relative normalizations of the coupling $g$ were chosen such that 
the beta function is the same for both the boson verses fermion cases.  
The S-matrix eq. (\ref{3dS})  here represents the scattering of two fermions of 
opposite spin.     Thus,  the fermion case also has a diverging scattering length
at the fixed point.   In the fermionic context,   the coupling $g_*$ is the boundary of
the so-called BEC/BCS crossover.    On the BCS side just above $g_*$,  the scattering length is negative,  which
implies effectively attractive interactions.    Here it is believed that at low enough temperatures there is a
phase transition to a strongly interacting version of superconductivity.     For $g< g_*$,   the scattering length is
positive,  signifying repulsive interactions,  and a bound state exists.    This bosonic bound state can undergo 
Bose-Einstein condensation,   hence this region is referred to as the BEC side.    
The physics is expected to be smooth as one crosses $g_*$.    In the  treatment of  the thermodynamics 
below,   we will work on the BCS side since here there is no need to incorporate a bound state into the
thermodynamics.

\section{Thermodynamics at the quantum critical point}

At the quantum critical point,  the only energy scales in the problem 
are the chemical potential $\mu$ and the temperature $T= 1/\beta$.   
This implies some
universal scaling forms for the various thermodynamic functions\cite{Ho}.
  The free energy density
has the form
\beq
\label{freeenergy}
\CF =  - \zeta (5/2) T  \, \lambda_T^{-3} \,  \, c(\mu/T) 
\eeq
where $\zeta$ is Riemann's zeta function and $\lambda_T = \sqrt{2 \pi/mT}$  
is the thermal wavelength.   
The  scaling function $c$ is only a function of $x\equiv \mu/T$.
 With the above normalization,  a single free boson has 
 $c=1$ in the limit of $x \to 0$.   
 It is also convenient to define the  scaling function $q$, which is a measure
of the quantum degeneracy, in terms of the density $n$    as follows: 
\beq
\label{nhat}
n \lambda_T^3 = q
\eeq
The two scaling functions $c$ and $q$ are of course related since
$n= - \d \CF / \d \mu$,  which leads to 
$q = \zeta (5/2) c'$,
where $c'$ is the derivative of $c$ with respect to $x$.   
Henceforth $b'$ will always denote the derivative of  a function $b (x) $ with respect to $x$.

The approach to the statistical mechanics of particles developed in complete generality in \cite{PyeTon} is 
based on the S-matrix.    On starts from a formula derived in \cite{Ma} for the partition function:
\beq
\label{ZS}  
Z = Z_0 +  \inv{4 \pi i}  \int  dE  e^{-\beta E}  \,  {\rm Tr}   \, {\rm Im}   \d_E  \log \hat{S} (E)  
\eeq
where $Z_0$ is the partition function for the free theory and $\hat{S} (E)$  is the off-shell 
S-matrix operator in the usual scattering theory.    Though the above expression is simple enough,
a considerable amount of additional work is required to turn it into something useful.   
The trace is over the multi-particle Fock space,   thus the above expression contains contributions
from N-body processes for all N.    For integrable theories in one spatial dimension,  the N-body
S-matrix factorizes into 2-body S-matrices,  and though it has never been proven,   the above expression
should recover the thermodynamic Bethe ansatz\cite{YangYang}.    

As described in detail in \cite{PyeTon},  one can develop a diagrammatic description of the
contributions to eq. (\ref{ZS}),  where vertices with 2N legs correspond to the logarithm of the S-matrix
for N particle scattering and lines connection the vertices are occupation numbers.    The free energy
can be obtained from a variational principle based on a Legendre transformation between the chemical
potential and the density.    The variational principle leads to an integral equation satisfied by
the occupation numbers with kernels involving the logarithm of the S-matrix.     In the present context,   we 
restrict ourselves to 2-body processes only,  which,  as explained in the last section,  can be 
calculated exactly.    This should be a good approximation if the gas is 
not too dense.    We also remark that this method is quite different than the  method used for example by
Nozieres-Schmitt-Rink\cite{Nozieres},   although both involve the S, or t-matrix.    The latter formalism does
not lead to the self-consistent integral equations we describe below,  but rather a BCS-like equation.  
We point out that this latter method was extended to the unitary gas in \cite{Koetsier}.

Let us now describe the main result derived in \cite{PyeTon}.    
  Consider again  for simplicity a single  component bosonic or fermionic 
gas.   The filling fractions,  or 
occupation numbers,  are parameterized in terms of a pseudo-energy
$\vep (\kvec )$:
\beq
\label{fill}
f(\kvec )  =  \inv{ e^{\beta \vep (\kvec ) } -s }
\eeq
which determine the density:
\beq
\label{dens}
n = \int \frac{d^3 \kvec}{(2\pi)^3} ~ \inv{ e^{\beta \vep  (\kvec ) } -s }
\eeq
where $s=1,-1$ corresponds to bosons, fermions respectively.
The consistent summation of 2-body scattering leads to 
an integral equation for the 
 pseudo-energy $\vep (\kvec)$,   analogous to the Yang-Yang integral equation.      It is convenient to define the
quantity:
\beq
\label{ydef}
y (\kvec ) = e^{-\beta (\vep(\kvec) - \omega_\kvec + \mu )}
\eeq
where $\omega_\kvec = \kvec^2 / 2m$.   Then $y$ satisfies the integral
equation 
\beq
\label{ykappa}
y (\kvec ) =  1 + \beta   \int  \frac{d^3 \kvec'}{(2\pi)^3} \, 
G(\kvec - \kvec' )  \frac{y(\kvec' )^{-1}}{e^{\beta \vep (\kvec')} -s}
\eeq
The free energy density is then
\beq
\label{freefoam2}
\CF = -T  \int \frac{d^3 \kvec}{ (2\pi)^3} \[ 
- s  \log ( 1- s e^{-\beta \vep  } ) 
-\inv{2}  \frac{ (1-y^{-1} )}{e^{\beta \vep} -s}   \]  
\eeq
The kernel $G$  is related to the logarithm of the 2-body S-matrix of the last section,
and depends on the coupling $g$.   
In the unitary limit  $g \to g_*$  the kernel simplifies greatly since $S= -1$:  
\beq
\label{kernel}
G(\kvec - \kvec' ) =  \mp  \frac{8 \pi^2}{m |\kvec - \kvec'|} , 
\eeq 
where the $-$ sign corresponds to $g$ being just below the fixed point
$g_*$, where the scattering length $a\to +\infty$ on the BEC side,
whereas the $+$ sign corresponds to $a\to - \infty$ on the BCS side.  
As explained in the last section,  we will  work on the BCS side.

Finally comparing with the definitions above for the scaling functions $c, q$ one finds:
\beq
\label{nhatsc}
q (x)   =  \frac{2}{\sqrt\pi} \int_0^\infty  d\kappa  \sqrt{\kappa} 
  \frac{  y(\kappa) z }{e^\kappa - s  y(\kappa) z}  
\eeq
and 
\beq
\label{cscale}
c =  \frac{2}{\sqrt{\pi} \zeta (5/2) }   
\int_0^\infty  d\kappa \sqrt{\kappa}  \(  -s  \log \( 1- s z y(\kappa)
 e^{-\kappa} \) 
- \inv{2}  \frac{ z ( y(\kappa) - 1 ) }{e^\kappa - s zy(\kappa)} \)
\eeq
where $z = e^{\mu/T}$ is the fugacity and the dimensionless  integration variable is $\kappa = \kvec^2 / 2 m T$.  
The ideal, free  gas limit corresponds to $y=1$ 
where $q= s \Li_{3/2} (s z)$ and  $c= s \Li_{5/2} (sz)/ \zeta(5/2)$,
where $\Li$ is the polylogarithm.    The BEC critical point of the
ideal gas occurs at $\mu=0$, i.e. $q=\zeta(3/2)$.

Consider now two-component fermions with the action (\ref{fermionaction}).  
There are two pseudo-energies $\vep_{\up, \down}$ satisfying two coupled
integral equations.   Due to the SU(2) symmetry,  for equal chemical potentials 
$\vep_\up = \vep_\down$.    
However  the  available phase space  for 2-particle scattering is doubled and 
the kernels have an extra $1/2$:
\beq
\label{Gbosferm}
G_{\rm fermi} = \inv{2} G_{\rm bose} 
\eeq
Thus  the two-component fermion reduces to
two identical copies of the above 1-component expressions, with the
modification (\ref{Gbosferm}).

\section{Phase transitions in temperature}

In the present context of quantum statistical mechanics,   the quantum critical point at $g=g_*$ 
does not signify a phase transition.    However in the unitary limit at fixed $g=g_*$  phase 
transitions may occur in regions of temperature versus density.    
From the scaling form of the free energy,  such a phase transition must occur at
a fixed,  specific value of $x= x_c =  \mu/T$.     We will continue to refer to such phase transitions
as critical points.   
The simplest case is BEC in the free boson theory,   where the phase transition occurs 
at $x_c = 0$,   which can be translated into  the well-known relation between temperature and 
density at the critical point  $n_c \Lambda_T^3 =  \zeta (3/2) = 2.61$.

\subsection{Fermionic case}

The integral equation for $y(\kappa)$, eq. (\ref{ykappa}),  can be
solved numerically by iteration.
 One first substitutes 
$y_0 = 1$ on the right hand side and this gives the  approximation
$y_1$ for $y$.  One then substitutes $y_1$ on the right hand side
to generate $y_2$,  etc.   For regions of $z$ where there are no
critical points,  this procedure converges rapidly, and as little
as 5 iterations are needed.    For fermions,  as one approaches 
zero temperature, i.e. $x$ large and positive,   more iterations are needed
for convergence.  The following results are based on 50 iterations.  

When $z\ll 1$,   $y \approx 1$, and the properties of the free ideal 
gas are recovered, since the gas is very dilute.   There are solutions
to eq. (\ref{ykappa}) for all   $-\infty < x < \infty$.  ($x=\mu/T$).  
The scaling function $c$ at zero chemical potential is 
$c(0) = 1.76$  compared to the free fermion value
$c_{\rm free} = 2 - 2^{-1/2} = 1.29$, thus the interactions have
a significant effect.   (These are twice the 1-component values.)

\begin{figure}[htb] 
\begin{center}
\hspace{-15mm} 
\psfrag{x}{$x=\mu/T$}
\psfrag{y}{$C_V/N$}
\includegraphics[width=10cm]{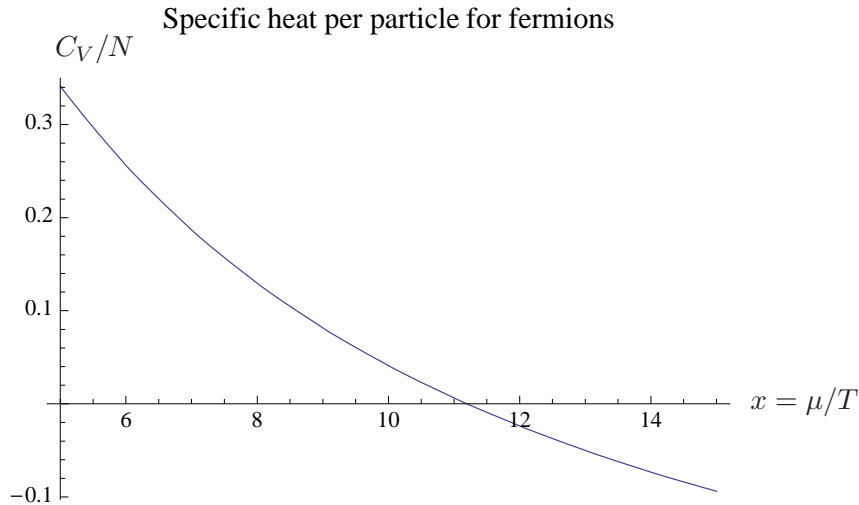} 
\end{center}
\caption{Specific heat per particle as a function of $x$ for fermions.}  
\vspace{-2mm}
\label{CVNF} 
\end{figure}

Whereas $c$ and $q$ are nearly featureless,  other quantities 
seem to indicate a phase transition at 
large density.    For instance,  the entropy per particle 
decreases with decreasing temperature up to $x < x_c \approx 11.2$.
 Beyond this point the entropy per particle
has the unphysical behavior of increasing with temperature.  
A further indication that the region $x>x_c$ is unphysical is
that the specific heat per particle becomes negative, as shown in
Figure \ref{CVNF}.  When $x\ll 0$,  $C_V/N$ approaches the 
classical value $3/2$.   This leads us to suggest a phase transition,
at $x=x_c =11.2$.  This value of $x$ can be expressed as
a critical temperature $T_c$ in units of the Fermi energy  $\ep_F = k_B T_F$.   From the 
definition $\ep_F = ( 3 \pi^2 n/\sqrt{2} )^{2/3} /m$,   one has
$T/T_F =  ( 4/3\sqrt{\pi} q )^{2/3} $. 
From the value $q(x_c)$, one finds 
  the critical temperature  
$ T_c /T_F  \approx 0.1$.  As we will show, our analysis of
the viscosity to entropy-density ratio suggests a higher $T_c/T_F$.  
There have been numerous estimates of $T_c/T_F$ based on various
approximation schemes, mainly using Monte Carlo methods on the lattice
\cite{Perali, LeeShafer,Wingate,Bulgac,Drummond,Burovski}, 
quoting results for $T_c/T_F$ between $0.05$ and $0.23$.  The work
\cite{LeeShafer} puts an upper bound $T_c / T_F < 0.14$,
and the  most recent results of Burovski et. al. quote $T_c/T_F =0.152(7)$.
Our result is thus consistent with previous work.

\subsection{Bosonic case}

The possibility of a phase transition in the unitary Bose gas is more subtle,
since a bosonic gas with attractive interactions is susceptible to mechanical collapse
to a denser state.    This issue has been studied for small negative scattering length in
a number of works\cite{Stoof,MuellerBaym,Thouless,YuLiLee},   and  the consensus is that mechanical collapse occurs
before BEC.    However it is not clear whether this conclusion can be extrapolated to
unitary limit where the scattering length is infinite.    In the approach described in the last section,
we found strong evidence for a phase transition to a strongly interacting version of BEC,  as we now explain.

\begin{figure}[htb] 
\begin{center}
\hspace{-15mm}
\psfrag{x}{$x=\mu/T$}
\psfrag{y}{$\vep (\kvec=0)/T$}
\psfrag{a}{$x_c$}
\includegraphics[width=10cm]{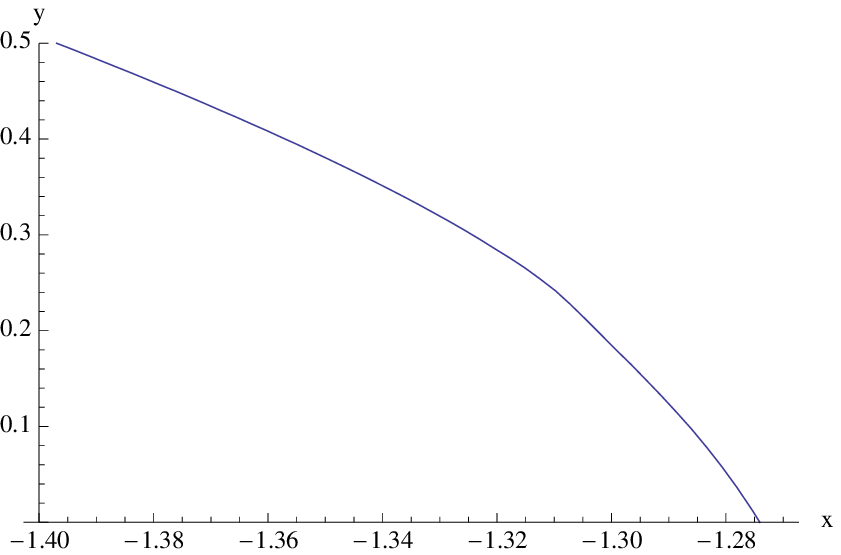} 
\end{center}
\caption{The pseudo-energy $\vep$ at $\kvec =0$  as a function of $x=\mu/T$.}  
\vspace{-2mm}
\label{epofx} 
\end{figure}

\begin{figure}[htb] 
\begin{center}
\hspace{-15mm}
\psfrag{x}{$\kappa =  \beta \kvec^2 /2m $}
\psfrag{y}{$f$}
\psfrag{a}{$x_c$}
\includegraphics[width=10cm]{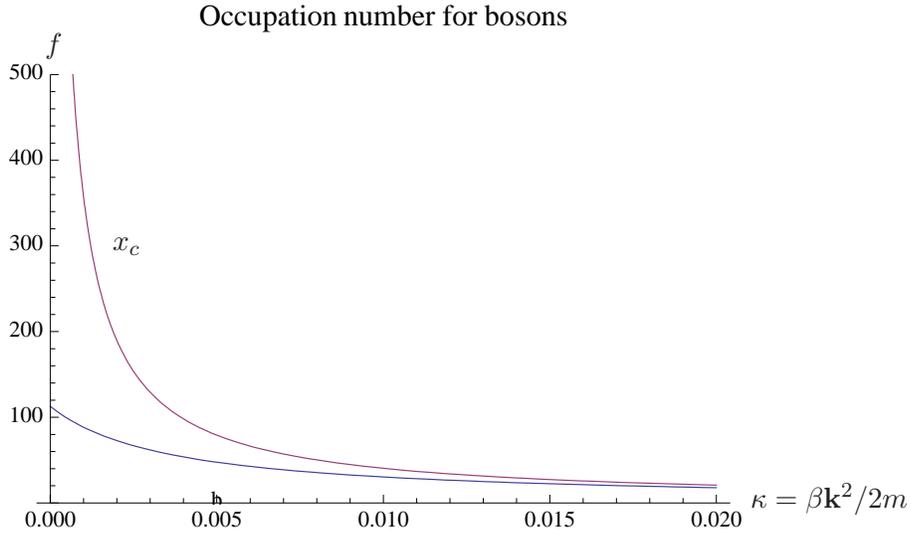} 
\end{center}
\caption{The occupation number $f(\kappa)$  for  $x=-1.275$ and $x_c = -1.2741$.}  
\vspace{-2mm}
\label{fB} 
\end{figure}

We again solved the integral equation (\ref{ykappa}) 
 by iteration, starting from $y=1$.   Since the occupation numbers decay
quickly as a function of $\kappa$,  we introduced a cut-off $\kappa < 10$.  
For $x$ less than approximately $-2$,  the gas behaves nearly classically.  
The main feature of the solution to the integral equation is that for
$x>x_c \equiv -1.2741$,  there is no solution that is smoothly connected to
the classical limit $x\to -\infty$.  Numerically,  when there is no solution
the iterative procedure fails to converge.  
 In Figure \ref{epofx}, we plot $\vep (\kvec=0 )$ as a function of
$x$,  and one sees that it goes to zero at $x_c$.   This implies the occupation
number $f$ diverges at $\kvec =0$ at this critical point.   
One clearly sees this behavior in Figure \ref{fB}.  
We also found that the compressibility diverges at $x_c$,  again consistent with BEC.

This strongly suggests  that there is a critical point at $x_c$ which  is a 
strongly interacting, scale invariant version of the ideal BEC.   
In terms of the density, the critical point is:
\beq
\label{xcb}
n_c \lambda_T^3  =  1.325, ~~~~~~~~~( \mu/T = x_c = -1.2741 )
\eeq
The negative value of the chemical potential is consistent with the
effectively attractive interactions.   The above should be compared with 
the ideal BEC of the free theory,  where $x_c = 0$ and 
$n_c \lambda_T^3 = \zeta (3/2) = 2.61$,  which is higher by a factor of 2.  

A  critical exponent $\nu$ characterizing the diverging  compressibility can be defined as 
\beq
\label{expnu}
\kappa  \sim  (T-T_c)^{-\nu}
\eeq
A log-log plot of the compressibility verses $T-T_c$ shows an approximately
straight line,   and we obtain   $\nu \approx 0.69$.    This should be compared with
BEC in an ideal gas,   where $\nu \approx 1.0$.     Clearly the unitary gas version 
of BEC  is  in a different universality class.

\section{Viscosity to entropy density  ratio}

Finally we turn to the intended focus of this article,  the ratio of the viscosity to entropy density.  
Consider first a single component gas.     The simplest expressions for the shear viscosity are based on
 kinetic theory,   where it   is related to the momentum transfer through 
an imaginary 2-dimensional plane cutting through the 3-dimensional bulk in the presence of a velocity 
gradient in the fluid flow.       It can be expressed as 
\beq
\label{shear.1}
\eta = \inv{3}  n  \vbar m \ell_{\rm free} 
\eeq
where $\vbar$ is the average speed and $\ell_{\rm free} $ is the mean
free path\cite{Reif}.     The mean free path is 
$\ell_{\rm free} = 1/(\sqrt{2} n \sigma)$ where $\sigma$ is the total
cross-section.   (The $\sqrt{2}$  comes from the ratio of the mean speed
to the mean relative speed.)  

Consider first the bosonic model.  (There were two errors by factors of 2 in the original discussion
 presented in \cite{PyeTonUnitary} which are here corrected.     However they turn out to  cancel and the main result below
eq. (\ref{shear.3}) is the same as before. )     In the unitary limit where 
the S-matrix $S=-1$,   the amplitude  is \cite{PyeTonUnitary}: 
\beq
\label{crossM}
\CM =  \frac{16 \pi i}{m |\kvec - \kvec'|}
\eeq
The cross section was calculated by  the standard manner in which it is related to 
an integration over final state momenta  and the square of the amplitude $\CM$ 
\beq
\label{shear.2}
\sigma =  \frac{ m^2 |\CM|^2}{8\pi}  
=  \frac{32 \pi}{|\kvec - \kvec' |^2}
\eeq
Finally,   
\beq \( |\kvec - \kvec'|^2 \) _{\rm ave}  =\(  \kvec^2  + \kvec'^2 -2 \kvec \cdot \kvec'  \) _{\rm ave}  = 
 2 m^2 \vbar^2 
 \eeq
 since the cosine of the angle between $\kvec$ and $\kvec'$ averages to zero.   
This gives   
\beq
\label{shear.3}
\eta_{\rm boson}  =  \frac{m^3 \vbar^3}{48 \sqrt{2} \pi} 
\eeq

For a unitary gas,  the relation between the energy, volume and pressure is the
same as for a free theory:  $E/V = 3 p/2$\cite{Ho}.  
Since the  pressure
is due to the kinetic energy, this implies
\beq
\label{shear.4}
\inv{2} m \vbar^2  =  E/N  =  \frac{3}{2} \frac{c}{c'} T   
\eeq
Thus $\eta \propto ( c T/c' )^{3/2}$.   In the high temperature limit,   the function $y\to 1$
and $c \propto Li_{5/2} (e^{x}) \propto x  $ and $c' \propto Li_{3/2} (e^x)  \propto x $ as
$x\to 0$.   Thus,  as expected $\eta \propto T^{3/2}$.
Since the  entropy density $s= - \d \CF / \d T$, one finally has
\beq
\label{etas}
\frac{\eta}{s} =  \frac{\sqrt{3\pi}}{8 \zeta (5/2)} 
 \(  \frac{c}{c'} \)^{3/2}  \inv{ 5 c /2 - x c' } 
\eeq

For two-component fermions,  the available phase space,  referred to as $\CI$ in 
\cite{PyeTonUnitary},   is doubled.
Also,  spin up particles only scatter with spin down.  This implies
the cross section  is $8$ times  smaller than  the  expression (\ref{shear.2}).   Since the entropy density is doubled  due to the two components,
this implies that $\eta/s$ is $4$ times the expression eq. (\ref{etas}),  where 
$c$ is the 1-component value appropriate to fermions.

\begin{figure}[htb] 
\begin{center}
\hspace{-15mm} 
\psfrag{X}{$T/T_F$}
\psfrag{Y}{$\eta / s$}
\includegraphics[width=10cm]{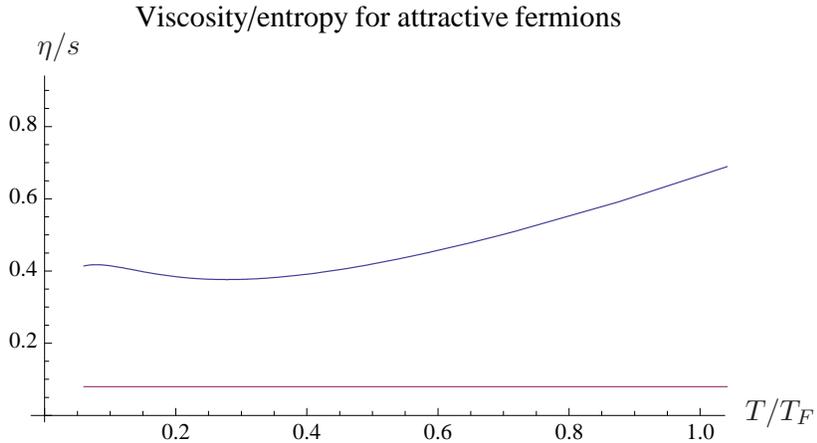} 
\end{center}
\caption{The viscosity to entropy-density ratio  as a function of
$T/T_F$ for fermions.  The horizontal line is $1/4\pi$.}  
\vspace{-2mm}
\label{etasF} 
\end{figure}

The ratio $\eta/s$ for fermions as a function of $T/T_F$ is shown in
Figure \ref{etasF},  and is in good agreement both quantitatively and
qualitatively  with the experimental data 
summarized in \cite{Schafer}.   The lowest value occurs at $x=2.33$, 
which corresponds to $T/T_F = 0.28$, and 
\beq
\label{etasFlim}
\frac{\eta}{s} > 4.72 \frac{\hbar}{4 \pi k_B}
\eeq
This is consistent with the  most recent  experimental data which shows 
 a minimum that is about $4-5$ times the bound\cite{Thomas}.      
Other,  considerably  more complicated  theoretical approaches,  e.g. using a
Kubo formula for the viscosity,  gives results around $6-7$ times the bound\cite{Enss}.

\begin{figure}[htb] 
\begin{center}
\hspace{-15mm} 
\psfrag{x}{$T/T_c$}
\psfrag{y}{$\eta / s$}
\includegraphics[width=10cm]{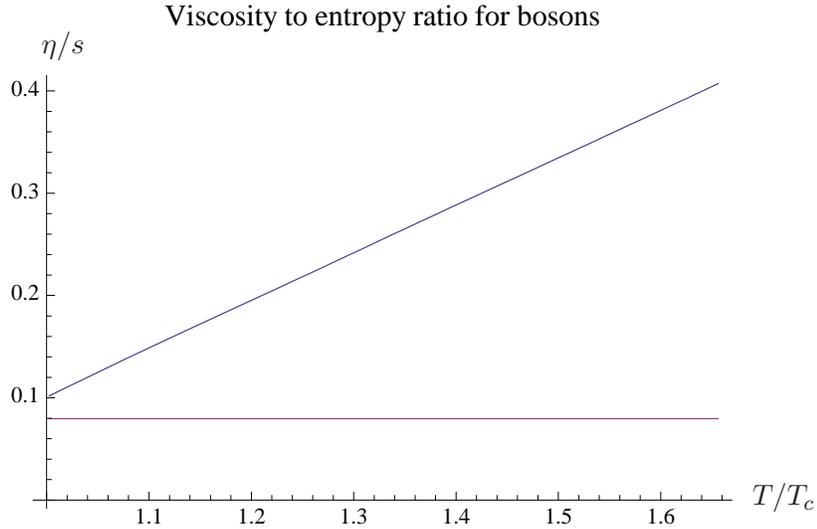} 
\end{center}
\caption{The viscosity to entropy-density ratio  as a function of
$T/T_c$ for bosons.  The horizontal line is $1/4\pi$.}  
\vspace{-2mm}
\label{etasB} 
\end{figure}

For bosons, the ration $\eta/s$ is plotted in Figure \ref{etasB} as
a function of $T/T_c$.   One sees that it has a minimum at the critical point,
where 
\beq
\label{etasBlim}
\frac{\eta}{s} > 1.26  \frac{\hbar}{4 \pi k_B}
\eeq
   Thus the bosonic gas 
at the unitary critical point is a more perfect fluid than that of fermions.

\section{Conclusions}

In summary,  we have calculated the ratio of viscosity to entropy density for
both Bose and Fermi gases in the unitary limit using the S-matrix based approach
in \cite{PyeTonUnitary}.     For the fermionic gas,  we found $\eta/s \geq $ is greater than
$4.7$ times the conjectured lower bound of $1/4\pi$,   which is  lower than the result of
other theoretical approaches,  and more consistent with the most recent experimental results\cite{Thomas}.   
We provided evidence that the unitary Bose gas is stable and has a strongly interacting  BEC phase transition. 
The same calculation indicates that the Bose case is a more perfect fluid,  with $\eta/s \geq 1.7$ times
the lower bound.

\section{Acknowledgments}

We wish to thank the editors of this special issue for the invitation to
present these results. 
   This work is supported by the National Science Foundation
under grant number  NSF-PHY-0757868.

\end{document}